\documentclass[12pt]{iopart}

\usepackage{textcomp}
\usepackage{setspace}
\usepackage{tipa}
\usepackage{verbatim} 
\usepackage{bbold}
\usepackage{amssymb}
\usepackage{iopams}
\usepackage{accents}
\usepackage{amscd}
\usepackage{graphicx}
\usepackage{epstopdf}
\usepackage{caption}
\usepackage{subfig}
\usepackage{pdfpages}

\def\aa{\hat{a}}
\def\ad{\hat{a}^{\dag}}

\def\AA{\hat{A}}
\def\AD{\hat{A}^{\dag}}

\def\be{\begin{equation}}
\def\ee{\end{equation}}
\def\bea{\begin{eqnarray}}
\def\eea{\end{eqnarray}}
\def\nn{\nonumber}
\def\eta{{\it et. al.}}

\definecolor{darkgreen}{rgb}{0,0.4,0}

\newcommand{\upb}{Integrated Quantum Optics, Universit\"at Paderborn, Warburger Strasse 100, 33098 Paderborn, Germany}
\newcommand{\prague}{FNSPE, Czech Technical University in Prague, Br\^ehov\'{a} 7, 119 15, Praha 1, Czech Republic}

\date{\today}

\begin{document}
\title{Driven Discrete Time Quantum Walks}

\author{Craig S.~Hamilton$^1$, Sonja Barkhofen$^2$, Linda Sansoni$^2$, Igor~Jex$^1$ and Christine Silberhorn$^2$}
\address{$ö1$ \prague}
\address{$ö2$ \upb}
\ead{hamilcra@fjfi.cvut.cz}

\begin{abstract}

We introduce the driven discrete time quantum walk, where walkers are added during the walk instead of only at the beginning. This leads to interference in walker number and very different dynamics when compared to the original quantum walk. These dynamics have two regimes, which we illustrate using the one-dimensional line. Then, we explore a search application which has certain advantages over current search protocols, namely that it does not require a complicated initial state nor a specific measurement time to observe the marked state. Finally, we describe a potential experimental implementation using existing technology. 

\end{abstract}



\maketitle

\section{Introduction}

Quantum walks (QW) are the quantum extension of the classical random walk with a quantum walker replacing the classical walker \cite{Kempe:2003p1645, VenegasAndraca:2012p8540}. When compared to the classical case they exhibit very different properties due to the interference of the walker with itself. This is ubiquitously manifested in the faster spreading of the walker (ballistically) throughout a topology (i.e. 1D line, 2D lattice) when compared to a classical walker which moves diffusely. A  QW on the one-dimensional line with the walker initially localised at a central position evolves in time to a probability distribution with two lobes moving ballistically away from the origin, whereas a classical walker produces a Gaussian probability distribution, which is characteristic of a diffusion process. 

There has been various theoretical and experimental progress in QWs. Theoretically QWs have been shown to be universal for quantum computation \cite{Childs:2009p1332, Lovett:2010p2308}, forming an implementation of the Grover search algorithm \cite{Ambainis:2004p1943} and other search algorithms \cite{Shenvi:2003p5446}. They have also been used to model transport processes such as quantum heat transport \cite{Asadian:2013p10070} and percolation \cite{Elster:2015p12200}. Various experimental implementations have also been explored in optical delay loops \cite{Schreiber:2010p5461, Schreiber:2012p5592, Hamilton:2011p4018}, trapped atoms \cite{Karski:2009p2506}, microwave scattering \cite{Bohm:2015p12316} and waveguide arrays \cite{Perets:2008p1878, Peruzzo:2010p3875, Owens:2011p4226, Sansoni:2012p5030}. 

QWs come in two different forms; continuous-time and discrete-time quantum walk (CTQW \& DTQW) \cite{Aharonov:1993p1091, Meyer:1996p2034}. CTQWs are described by a Hamiltonian, such as a collection of coupled oscillators, and evolve under the Schr\"{o}dinger equation. In DTQWs the walkers have an internal state, called a coin state, and evolve iteratively by the discrete application of two unitary operations, the coin operator and the step operator. We recently described a new type of continuous time quantum walk where walkers can be coherently created and destroyed throughout the walk \cite{Hamilton:2014p10181}, which we termed a driven QW. This work was motivated by a new experimental realisation of QW in waveguide arrays with a nonlinear down-conversion process \cite{Kruse:2013p8361} and is similar to previous work where walkers were incoherently added at each time step \cite{Regensburger:2011p4911}.

In this paper we extend our work on driven QWs and introduce a new type of quantum walk, the driven discrete-time quantum walk (DDTQW), where extra walkers are added during the walk. This paper is structured as follows: In the next section we describe the theory of DTQW. Following that we introduce the new type of quantum walk, called a driven quantum walk, in the discrete-time formalism. Next we explore a driven QW version of the Grover search algorithm and its advantages over the original algorithm. Finally we look at possible experimental implementations of this idea. 

\section{Theory of DTQW}

A DTQW takes place on a topology of connected vertices, typically a 1D-line or a 2D lattice. The walker(s) that moves on this topology, called the position space $\{|x\rangle\}$, has an internal state, the coin space. The walker evolves iteratively by applying the coin operator, $\hat{C}$, (acting solely upon the coin space) that `flips' the coin, and then a step operator, $\hat{S}$, that moves the walker in a direction that depends upon the state of the coin. The walker's wavefunction at time $t$ is described by $|\phi\rangle_t$, which evolves according to,
\be \label{eq:QW_unitary}
|\phi\rangle_{t+1} = \hat{S} \hat{C} |\phi\rangle_t.
\ee
and can be written as,
\be
|\phi\rangle_t = \sum_{c,x} a_{c,x,t} | C \rangle \otimes |x\rangle \label{DDTQW_state}
\ee
where $| C\rangle$ is the coin space, $|x\rangle$ is the position space and $a_{c,x,t}$ are complex coefficients.
We term the basis that (\ref{DDTQW_state}) is written in the `physical basis' and this is the basis in which measurements typically are performed. At the end of the walk (t=T) the probability distribution of the walker's position, $P(x)=\sum_c |a_{c,x,T}|^2$, is the main figure of interest and shows different dynamics than the classical random walk \cite{Kempe:2003p1645, VenegasAndraca:2012p8540}.

In this paper we will also describe the QW in the eigenbasis of the operators, as this allows for easy analytical calculations. This eigenbasis diagonalizes the above dynamics of the coin and position space, and the evolution operator is now:
\be
\hat{E} = \sum^N_{j=1} \hat{E}_j = \hat{T} \hat{S}\hat{C} \hat{T}^\dag,
\ee
where $\{\hat{E}_j\}$ are the set of eigenoperators and $\hat{T}$ is the transformation that takes the state from the physical basis $\left\{|C\rangle|x\rangle\right\}$ to the eigenbasis $\left\{|E_j\rangle\right\}$
\be
|\Phi_E\rangle_t =\hat{T} |\phi\rangle_t=\sum_jb_{j,t}|E_j\rangle.
\ee
The eigenvectors $|E_j\rangle$ of $\hat{S}\hat{C}$ form the columns of $\hat{T}$ with eigenvalues $\{e^{i\,\omega_j}\}$, $\omega_j$ being termed the eigenfrequencies. The eigenstates are combinations of the coin and position states,
\be
|E_j \rangle = \sum_{c,x} T_{j,c,x} |C\rangle\otimes |x\rangle,
\ee
where the number of eigenstates is $N=|C|\times |x|$ ($|\,.\,|$ is the size of that state space). The dynamics of the original quantum walk become straightforward to solve in the eigenmode basis,
\be
|\Phi_E\rangle_{t+1} = \hat{E} |\Phi_E\rangle_t = \hat{E} \sum_j b_{j, t} |E_j\rangle = \sum_j b_{j ,t} e^{i \omega_j} |E_j\rangle=\sum_j b_{j ,0} e^{i \omega_j(t+1)} |E_j\rangle.
\ee

\section{Driven DTQW}

Now we extend the DTQW to include the possibility to create and destroy walkers during the walk. We refer to this case as driven DTQW. In this description we do not have a definite number of walkers but use instead the average number of walkers, i.e. walker intensity, as our alternative measure of walker number. We assume our walkers to be indistinguishable bosons and will use the bosonic annihilation and creation operators in describing the system dynamics. Here $\ad_{c,x}$ is the creation operator for a photon (walker) with coin state `$c$' at position `$x$'. The action of the coin ($\hat{C}$) on  $\aa,\ad$ is,
\be
\hat{C} \left ( \begin{array}{c} \hat{a}_{R,x} \\ \hat{a}_{L,x} \end{array}\right)\hat{C}^{\dagger}= \left ( \begin{array}{cc} C_{R,R} & C_{L,R} \\ C_{R,L} & C_{L,L} \end{array}\right)
\left ( \begin{array}{c} \hat{a}_{R,x} \\ \hat{a}_{L,x} \end{array}\right)
\ee
where the $C_{ij}$ are elements of the coin operator. The coin operator could also be position dependent ($\hat{C}_x$), which we will make use of later in the paper. The action of the step operator is,
\be
\hat{S}\,\ad_{R,x}\hat{S}^\dag =  \ad_{R, x+1}  ;\,\,\, \hat{S}\,\ad_{L,x}\hat{S}^\dag = \ad_{L, x-1}.   
\ee
These operators are Gaussian operations and can be generated from multimode beamsplitter Hamiltonians \cite{Furusawa_VanLoock} that describe the original passive DTQW. Eigenoperators for creation and annihilation of photons can also be defined, 
\be
\AD_j = \sum_{c,x} T^j_{c,x} \ad_{c,x},
\ee
and the action of the operator $\hat{E}$ on them is given by
\be
\hat{E}^{}_j \AD_j \hat{E}^\dag_j = \AD_j e^{i\omega_j}.
\ee

In order to generate our walker creation/destruction we will use the displacement operator,
\be
\hat{D}(\alpha_{c,x}) = \exp \left [ \alpha^{}_{c,x} \,\ad_{c,x} - \alpha_{c,x}^*\,  \aa^{}_{c,x} \right ],
\ee
where $\alpha_{c,x}$ is the amplitude of the displacement operator for a particular mode $\{c,x\}$. For multiple modes the total displacement operator is simply the product of the individual operators. The dynamics of the original DTQW (after $t$ steps) in the physical basis can then be written with the inclusion of the displacement operator as,
\be
|\phi\rangle_t = (\hat{S} \hat{C} )^t \prod_j \hat{D}(\alpha_j) |0\rangle_{t=0}, \label{ori_DTQW}
\ee
with $j=\{c,x\}$. In typical experiments the initial $\hat{D}(\alpha_j)$ is localized on a single position, i.e. $x=0$, with a certain coin state i.e. $j=\{[R,0], [L,0]\}$.

In this paper we study the new walk, which we term a driven DTQW, where we now inject walkers at each time-step of the walk, instead of just at the start of the walk. In this case the wavefunction evolves according to,
\be
|\phi\rangle_t= \prod_{k=1}^t\left  (\hat{S} \hat{C} \prod^N_{j=1} \hat{D}( \alpha_{j,k}) \right ) |0\rangle_{t=0}. \label{DDTQW_equ}
\ee
The $\alpha_j$ will determine which eigenmodes the pump couples to and the time dependence decides which eigenmode is phase-matched, i.e. an eigenmode where the walkers accumulate in time. At the end of the walk dynamics we will be interested in  the intensity of walker distribution, $I_{c,x} = \langle \ad_{c,x}\aa_{c,x} \rangle$, as we no longer have a single walker in our system. In subsequent sections we will calculate the intensity of walkers in both the physical basis and the eigenmode basis to demonstrate the dynamics of the walk. 

We now wish to compare the DDTQW, (\ref{DDTQW_equ}), to the traditional quantum walk by the re-ordering of the operators $\hat{S} \hat{C}$ and $\hat{D}$ to resemble the original DTQW, (\ref{ori_DTQW}), i.e. state creation followed by the walk operators.

\subsection{Comparison to original walk}

Our method to compare the two walks relies on re-arranging the evolution operators of the walk to order them in such a way that all the displacement operators, $\hat{D}$, are to the right of the walk operators, $\hat{S} \hat{C}$. To do this we analyze the walk in the eigenbasis which makes solving the dynamics simple, as we will discuss in the following section. 

The eigenoperator $\hat{E}_j$ can be written using the operators $\AD_j$ and $\AA_j$ as
\be
\hat{E}_j  = \exp \left [ -i \omega_j \AD_j \AA^{}_j \right]
\ee
The displacement operator $\hat{D}(\alpha_{c,x})$ for the individual, physical sites $\{c,x\}$ will be transformed to the displacement operator for the eigenmodes, $\hat{D}_E(\alpha_j)$. A single step for the driven QW in the eigenmode basis is now,
\be
|\Phi_E\rangle_{t+1} = \hat{E} \, \prod^N_{j=1} \hat{D}_E(\alpha_{j,t})  |\Phi_E\rangle_t
\ee
where $|\Phi_E\rangle_t = \hat{T}|\phi \rangle_t $, the state in the eigenbasis. For $t$ steps, starting from the initial vacuum state, this is
\be
|\Phi_E\rangle_{t} = \prod_{k=1}^t \left (\hat{E} \, \prod_{j=1}^N \hat{D}(\alpha_{j,k})  \right)  |0\rangle_{t=0} \label{eigen_dyn}
\ee
We now wish to order expression (\ref{eigen_dyn}) such that all the displacement operators are to the right of the walk operators, i.e.
\be
(\hat{E} \hat{D})^t |0\rangle_{t=0} \rightarrow  \hat{E}^t \, \hat{D'}^t |0\rangle_{t=0} = \hat{E}^{\,t} |\Phi_E \rangle_{t=0} \rightarrow (\hat{S}\, \hat{C})^t |\phi \rangle_{t=0},
\label{eq:ord}
\ee
where $|\phi\rangle_0$ is the `initial' state generated by the action of the displacement operators on the vacuum state (in the physical basis). This now resembles the traditional quantum walk with an initial state of walkers propagating under the quantum walk operators $\hat{S} \, \hat{C}$. 

Using standard relations for the evolution of the annihilation/creation operator \cite{Barnett_Radmore}, we can write the following for the displacement operator, 
\be
\hat{D}_E(\alpha_j) e^{-i \omega_j \AD_j\AA_j} = e^{-i \omega_j \AD_j \AA_j} \hat{D}_E(\alpha_j e^{i \omega_j}), 
\ee
i.e. the displacement operator acquires a phase as it is re-ordered with the phase operator (different eigenmodes commute, acquiring no extra phase). To illustrate, 
\bea
\prod_{t=1}^3\left ( \hat{E} \, \hat{D}(\alpha_{t}) \right )&=& \hat{E} \, \hat{D}(\alpha_{t=3}) \,  \hat{E} \, \hat{D}(\alpha_{t=2}) \, \hat{E} \,  \hat{D}(\alpha_{t=1})  \nn \\
&=&\hat{E} \hat{D}(\alpha_{t=3}) \,  \hat{E}^2 \,  \hat{D}(e^{i \omega} \alpha_{t=2}) \,  \hat{D}(\alpha_{t=1}) \nn \\
&=&\hat{E}^3 \, \hat{D}(e^{i 2 \omega}\alpha_{t=3}) \,    \hat{D}(e^{i \omega} \alpha_{t=2}) \,  \hat{D}(\alpha_{t=1}) \nn \\
&=& \hat{E}^3 \, \hat{D}(\alpha_{F}),
\eea
where the product of the displacement operators will yield a final (yet to be determined) displacement operator, $\hat{D}(\alpha_{F})$. We have now re-ordered the expression so that it resembles (\ref{eq:ord}), the original DTQW, where state creation is followed by the QW dynamics. This `initial state' will in general be de-localized in physical space and therefore using a driven QW a wider class of states can be created and utilised in the walk dynamics. We are now interested in the product of the displacement operators, $\hat{D}(\alpha_{F})$, as they determine the dynamics of the walk. The amplitude of the $\hat{D}_E(\alpha_j)$ can be controlled by the pump shape in the physical basis, $\hat{D}(\alpha_{c,x})$. In general the $\alpha_{j,t}$ will have a step dependent phase that counter-acts the action of the phase operator and below we will show how this is crucial for the dynamics of the driven QW. 

\subsection{Dynamics of the DDTQW}\label{sec_phase_match}

We now focus on the dynamics of products of displacement operators for a single eigenmode,
\be
\prod^{N-1}_{t=0} \hat{D}(\alpha_j e^{i \omega_j t}) = \prod^{N-1}_{t=0} \hat{D}(|\alpha_j| e^{i \Delta_j t}),
\ee
where we assume a linear step dependent phase $\alpha_{j,t} = |\alpha_j| \exp(i \phi \, t)$, and we term $\Delta_j = \phi - \omega_j$ the phase-mismatch. The phase mismatch is important as it determines the dynamics of the driven QW. The intensity of eigenmode that is phase matched ($\Delta_j =0$) will grow quadratically in time, whereas phase mismatched modes ($\Delta_j \ne 0$) will oscillate between zero and a maximum intensity. We study these two situations below.

\subsubsection{Dynamics of phase-matched modes}

In this case the time-dependent phase added to the displacement operator would be equal to the eigenfrequency $\omega_j$ of the mode, yielding $\Delta = 0$. An example of phase-matched displacement operators is,
\be
\prod^{t-1}_{j=0} \hat{D}(\alpha) = \hat{D}(\alpha) ^t = \hat{D}(t \alpha ),
\ee
where the `in-phase' displacement operators add constructively and the intensity of walkers, in that eigenmode, grows quadratically, $I=t^2|\alpha|^2$, with the number of steps of the driven QW. 

\subsubsection{Non-phase matched modes}

Non-phase-matched modes, where $\Delta \ne 0$, have different dynamics,
\be
 \prod_{t=0}^{t-1} \hat{D} \left (|\alpha| e^{i \Delta t} \right ).
\ee
The final displacement operator after t steps is (for $P=e^{i \Delta} \ne \pm1$)
\be
\hat{D}\left(|\alpha| \frac{1-P^t}{1-P}\right) \exp \left [  i |\alpha|^2   \sum^{t-1}_{n=1} \sum^{n}_{k=1} \sin (k \Delta) \right  ].
\ee
The intensity of the mode oscillates in time with a frequency that depends upon the phase-mismatch,
\be
I = |\alpha|^2 \frac{\sin^2 ( t \Delta/2 )}{\sin^2 (\Delta/2)}.
\ee
In the next section we demonstrate these two regimes by simulating the dynamics of the driven DTQW on 1D topology. 

\section{Results on the 1D line}

In this section we illustrate the DDTQW by exploring the dynamics on a 1D line, where we plot the intensity of walkers at each physical vertex, and also the intensity of the walkers in the eigenbasis. 

The walker has two coin states, R(ight) and L(eft), with 5 vertices and reflecting coin operations at the two ends of the chain. The reflecting coins at the boundary vertices are, 
\be
\hat{\sigma}_x=\left ( \begin{array}{cc} 0 & 1 \\ 1 & 0 \end{array} \right ),
\ee 
and the middle three vertices have the Hadamard coin,
\be
\hat{H} = \frac{1}{\sqrt{2}}\left ( \begin{array}{cc} 1 & 1 \\ 1&-1 \end{array} \right ).
\ee
We then add walkers to the central vertex at every time step. The walkers that enter at the $t^{\mathrm{th}}$ step are in a coherent state, $|\alpha| e^{-i \omega_j \, t}$ (amplitude $|\alpha|$ and time dependent phase $\omega_j \, t$), with a single coin state, `R'. It is this frequency $\omega_j$ that we vary in our simulations, as it corresponds to the eigenfrequency of a particular eigenmode $j$ as described in the previous section. By varying the eigenfrequency different eigenmodes are phase-matched, creating photons in those eigenmodes and thus different physical modes for measurement. We plot the intensity in both bases throughout the walk. 

\begin{figure}[!tbp]
  \centering
  \subfloat{(a)}\includegraphics[width=0.4\columnwidth]{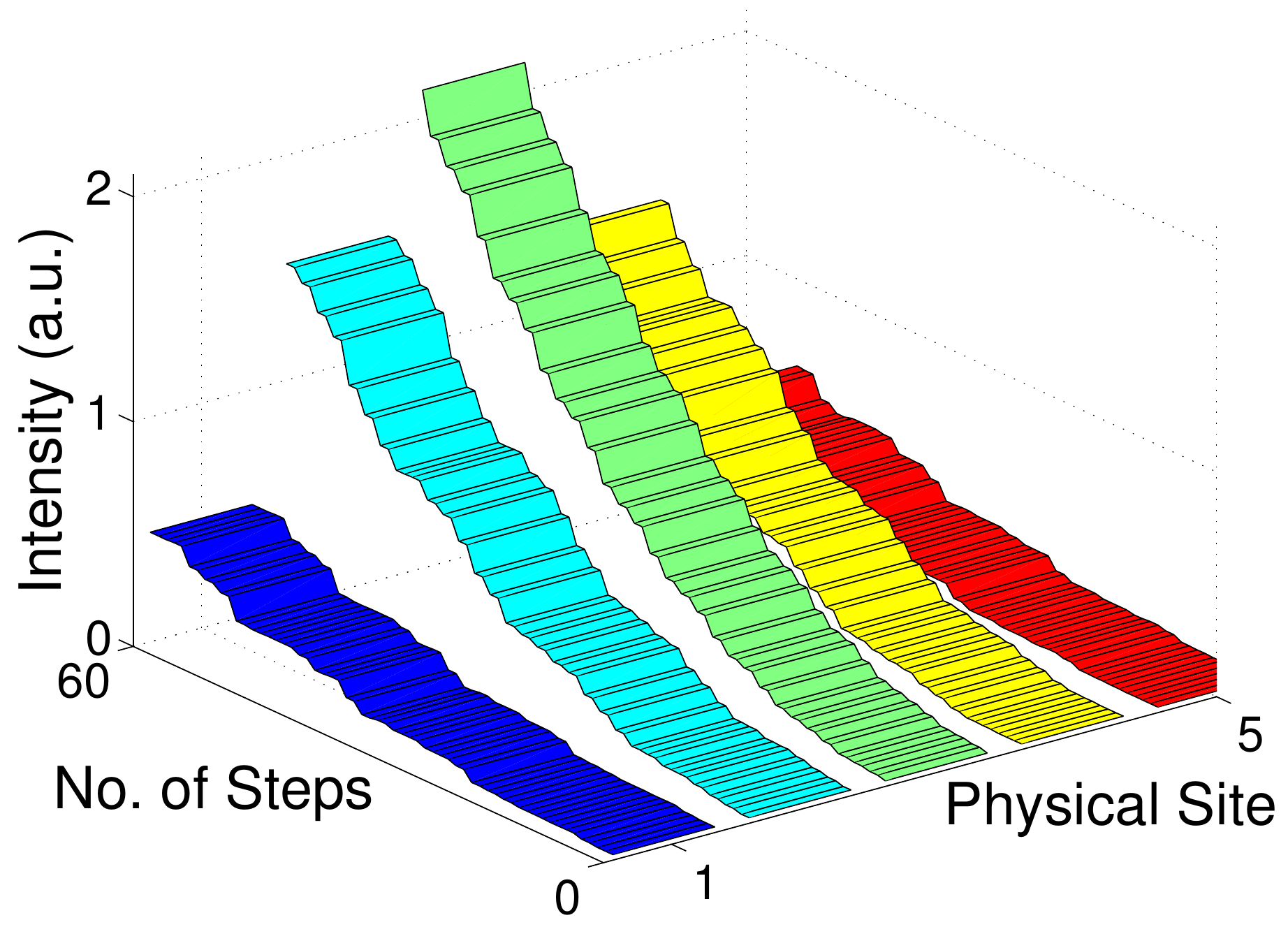}
  \hfill
  \subfloat{(b)}\includegraphics[width=0.4\columnwidth]{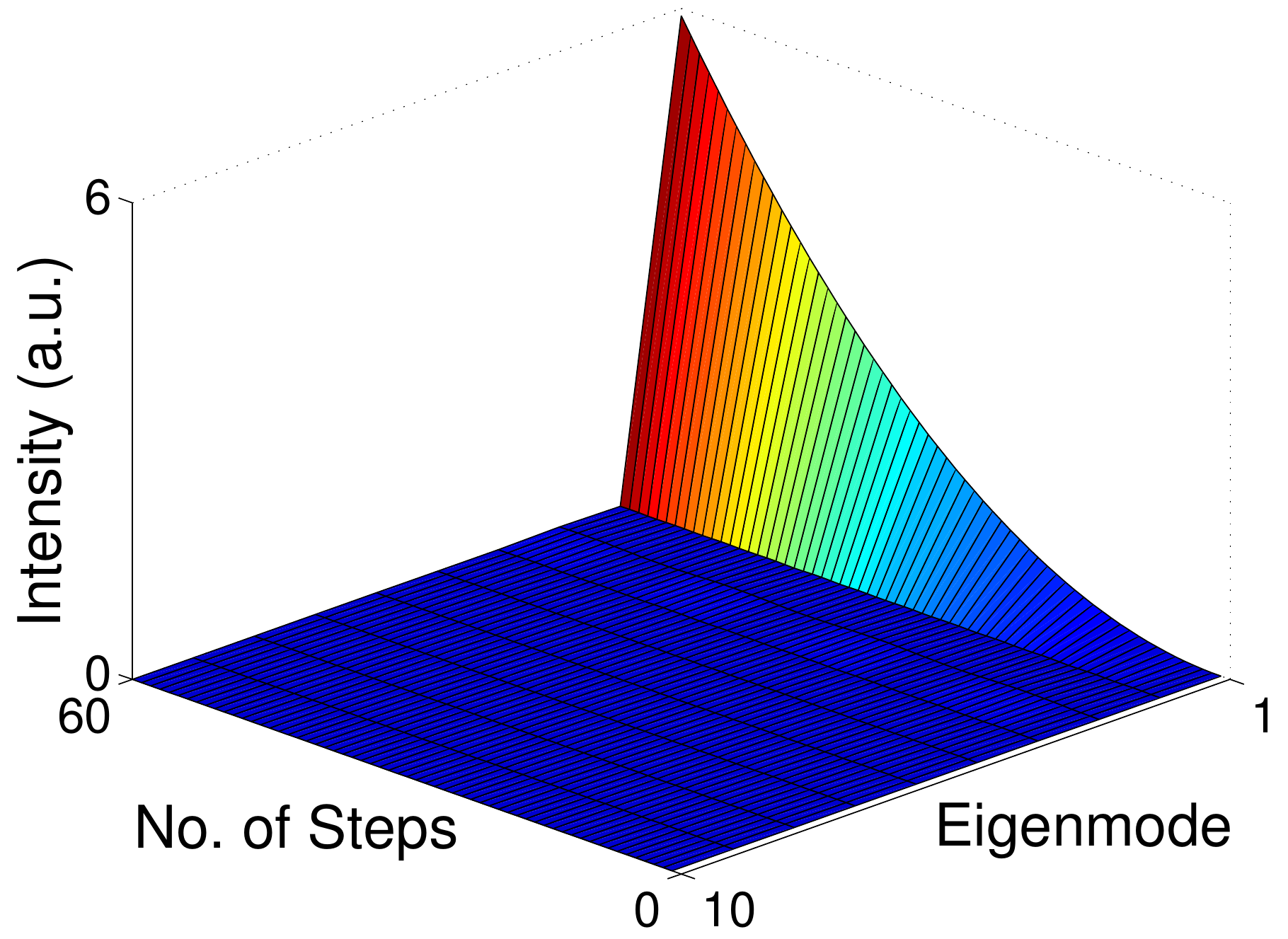}
\caption{Dynamics of the system in the position basis (coin space traced over)(a) and eigenmode basis (b)  when the $1^{st}$ eigenmode is phase-matched.}\label{fig_eg1}
\end{figure}

Figure~\ref{fig_eg1} shows the dynamics where the walkers are added to the central position with a phase that corresponds to a particular eigenfrequency, namely $j=1$. Figure \ref{fig_eg1}(a) shows the walker dynamics at each physical site (after tracing over the coin states), while figure \ref{fig_eg1}(b) shows the dynamics in the eigenmode basis. It can be seen clearly that only the phase-matched mode contributes to the dynamics, where, as expected, the intensity grows quadratically in time and the final photon number $\approx 6$ photons.

\begin{figure}[!tbp]
  \centering
  \subfloat{(a)}\includegraphics[width=0.4\columnwidth]{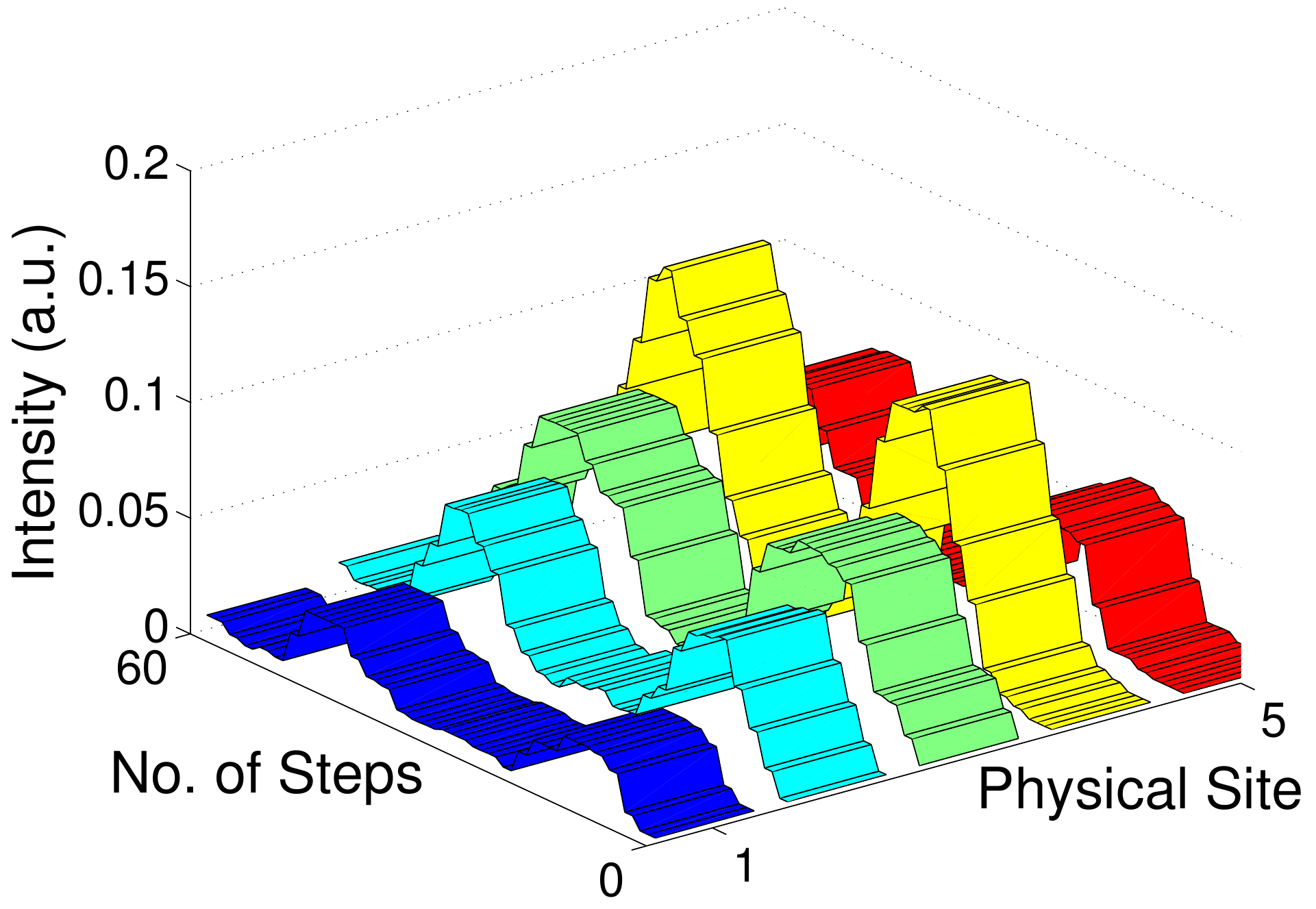}
  \hfill
  \subfloat{(b)}\includegraphics[width=0.4\columnwidth]{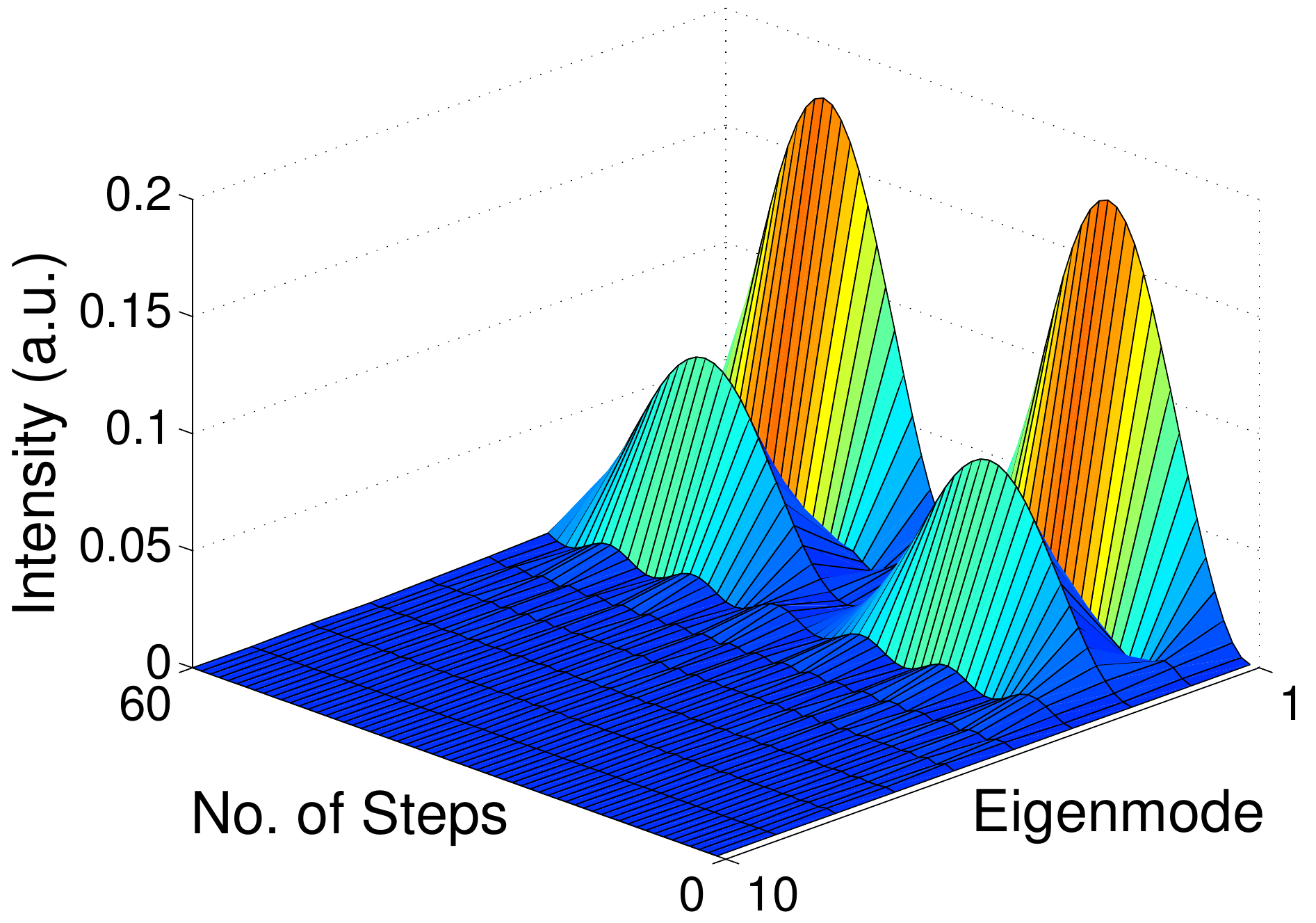}
\caption{Dynamics of the system in the position basis (coin space traced over)(a) and eigenmode basis (b)  when no eigenmode is phase-matched.}\label{fig_eg7}
\end{figure}

For figure~\ref{fig_eg7} we use the same walk topology with identical coin operations as before and we add the walkers in the same position. This time, however, we do not phase match to any eigenfrequency but at a frequency between the $1^{st}$ and $3^{rd}$ eigenfrequencies. The oscillations of the non-phase-matched dynamics can be clearly seen in both bases. The maximum intensity of photons in this regime, $\approx 0.3$ photons , is much lower than in the previous, phase-matched case.

These plots show how the dynamics of the walk depend upon the time-dependent phase of the walkers entering the walk. Different phases lead to the walkers accumulating in different eigenmodes of the system and in turn different physical modes of the system. In the next section we discuss an application of these phase-matched dynamics, a search protocol based on the Grover search algorithm. Following that we propose an experimental implementation of this one-dimensional line and a two-dimensional lattice using an existing optical time-delay loop architecture.

\section{Driven DTQW search}
      
In this section we apply the DDTQW idea to the QW version of the Grover search algorithm and construct a driven search algorithm.
We will start with the QW-scheme for Grover's search developed by Ambainis, Kempe and Rivosh (AKR) \cite{Ambainis:2004p1943}. The AKR scheme performs Grover's search on a DTQW with periodic boundary conditions, with the Grover matrix as the coin operator,

\be
\hat{C}_G=\frac{1}{2}\left [ \begin{array}{cccc}
-1 & 1 & 1 & 1\\
1 & -1 & 1 & 1\\
1 & 1 & -1 & 1\\
1 & 1 & 1 & -1
\end{array} \right ],
\ee
at every vertex except for the marked vertex, which has the identity coin $-\hat{\mbox{I}}_4$, and the step operator has an additional $\sigma_x$ term after it (called the flip-flop step operator by AKR). The initial state is a coherent, uniform distribution over all vertex and coin states which then evolves under the operator $\hat{U}=\hat{S}\hat{C}$. This evolves the extended state into one localised on the marked vertex in a time proportional to $\sqrt{N}$ ($\sqrt{N\, \mbox{ln} N}$for a 2D lattice), where $N$ is the total number of vertices of the lattice. 

Here we use an extension of the AKR scheme introduced by Hein and Tanner \cite{Hein:2009p2527, Hein:2010p9503} to propose a driven search algorithm for the QW Grover search. Hein and Tanner introduce $M$ positions on the d-dimensional lattice marked with the same coin ($-\mathbf{I}_4$) and analysed the eigenvectors and eigenvalues of this lattice. They show that the dynamics of this system can be reduced to a ($M$+1)-level system comprising a single extended state over all vertices and $M$ states that are localized on the marked vertices. The Hamiltonian of this reduced ($M$+1)-level system is given by an `arrowhead' matrix, which represents the topology of a star graph (a graph with a single central hub and $M$ `leaves' connected to this hub only) and its eigenvectors and eigenvalues can be given analytically. There is such an eigenvector that only has support on the leaves of the graph, has an eigenvalue of unity (an eigenfrequency $\omega =0$) and is independent of the position of the marked vertices. The other eigenvectors of interest in the system have frequencies that scale as $\sqrt{N\, \mbox{ln} N}$ (for a 2d lattice) and $\sqrt{N}$ for higher dimensions \cite{Hein:2009p2527, Hein:2010p9503} and are also independent of the marked vertices position.

\begin{figure}[phtb]
\begin{center}
\includegraphics[trim= 5cm 1cm 5cm 1cm , scale=0.45]{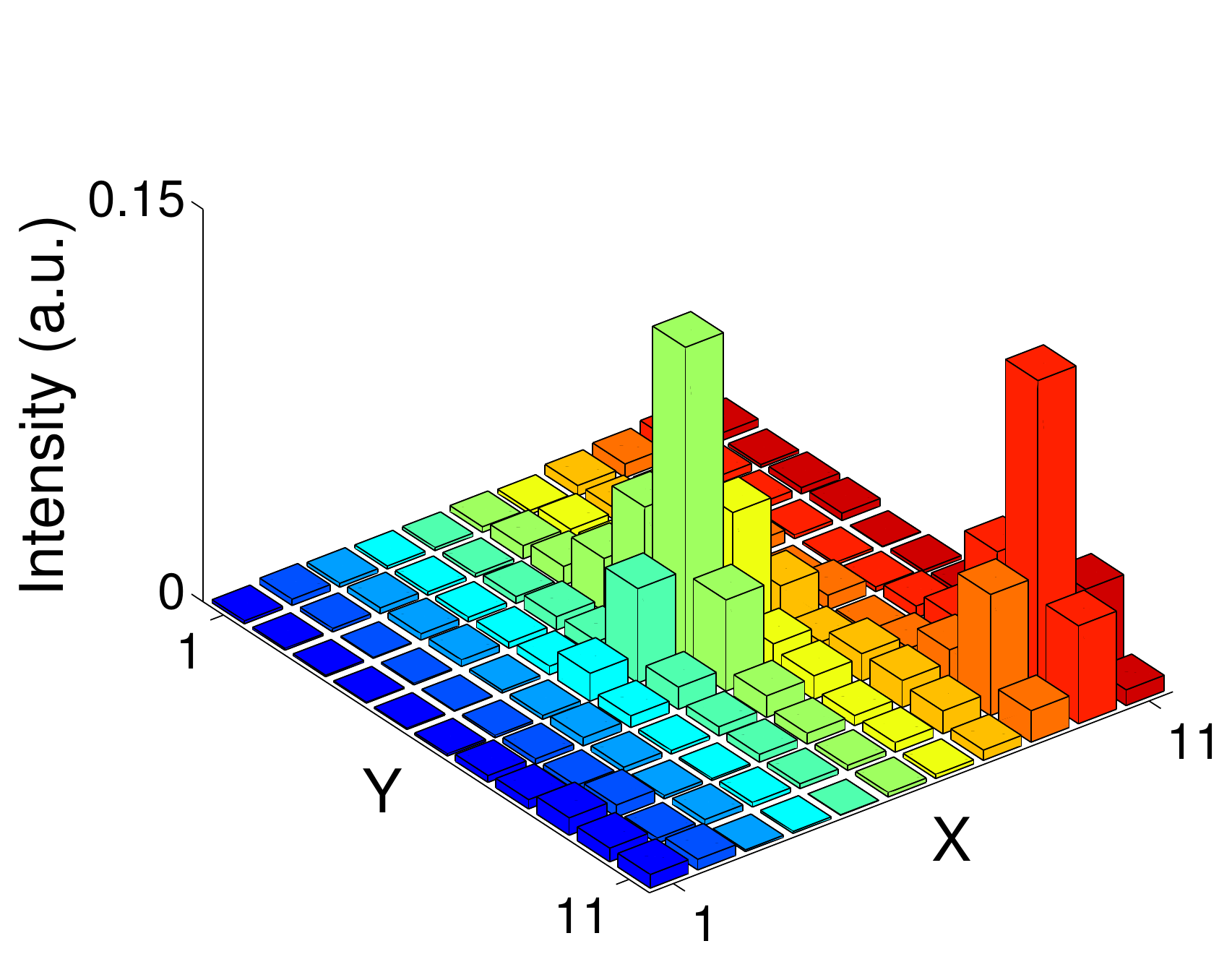}
\end{center}
\caption{Distribution of the eigenvector of interest in the physical basis. Coin states have been traced over.  }\label{eigen_vector_grover}
\end{figure}

\begin{figure}[phtb]
\begin{center}
\includegraphics[trim= 5cm 1cm 5cm 1cm , scale=0.35]{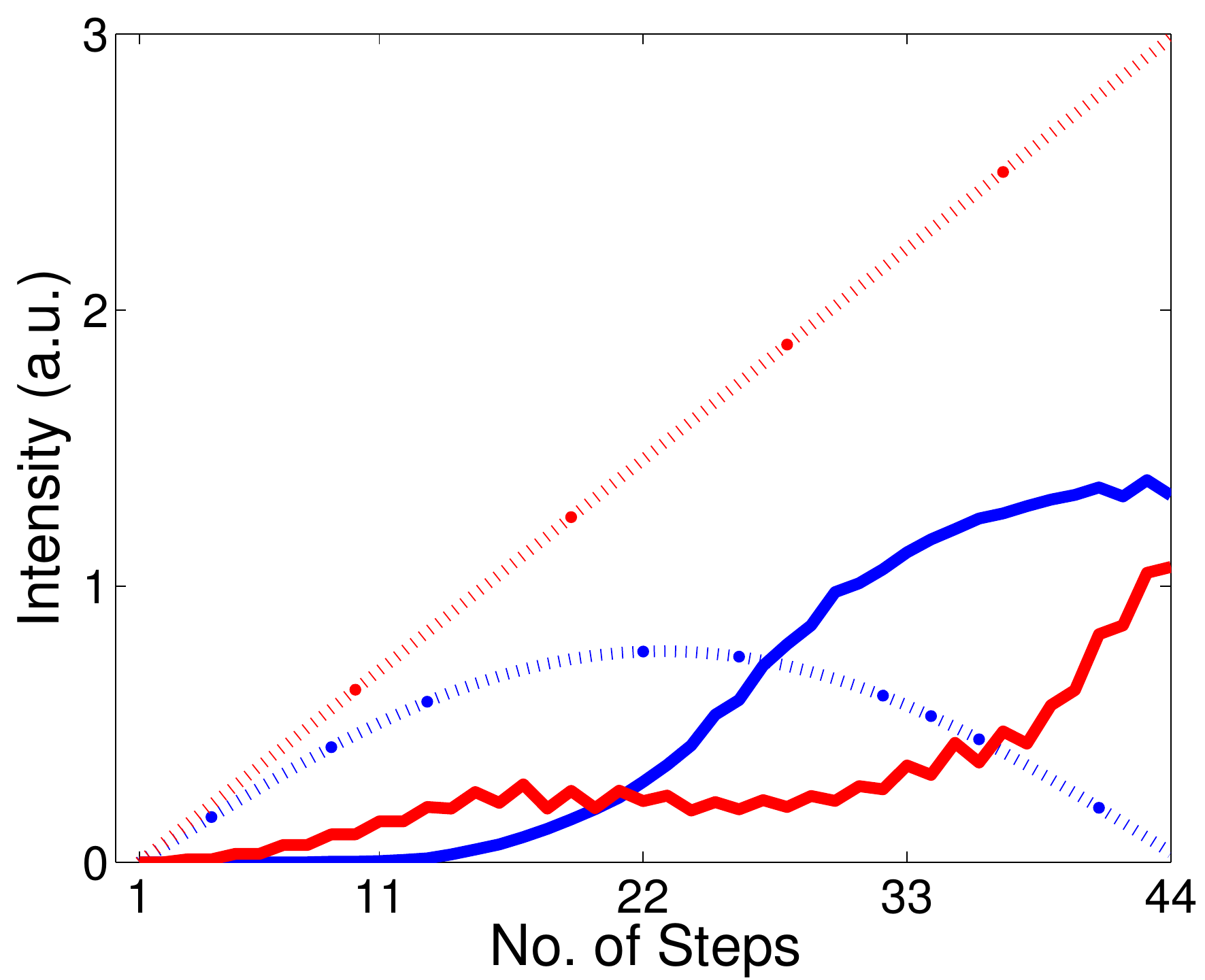}
\end{center}
\caption{Intensity of walkers vs. Time in the eigenmodes of interest and in the two physical sites of interest. Dotted lines: (Red) Phase-matched eigenmode that encodes the two marked vertices (shown in figure \ref{eigen_vector_grover}) (Blue) Two closest eigenmodes in eigenfrequency to zero. Solid lines: (Red) Walker Intensity at central vertex (Blue) Walker intensity at target vertex. }\label{ddtqw_grover_evo}
\end{figure}

\begin{figure}[phtb]
\begin{center}
\includegraphics[trim= 5cm 1cm 5cm 2cm , scale=0.45]{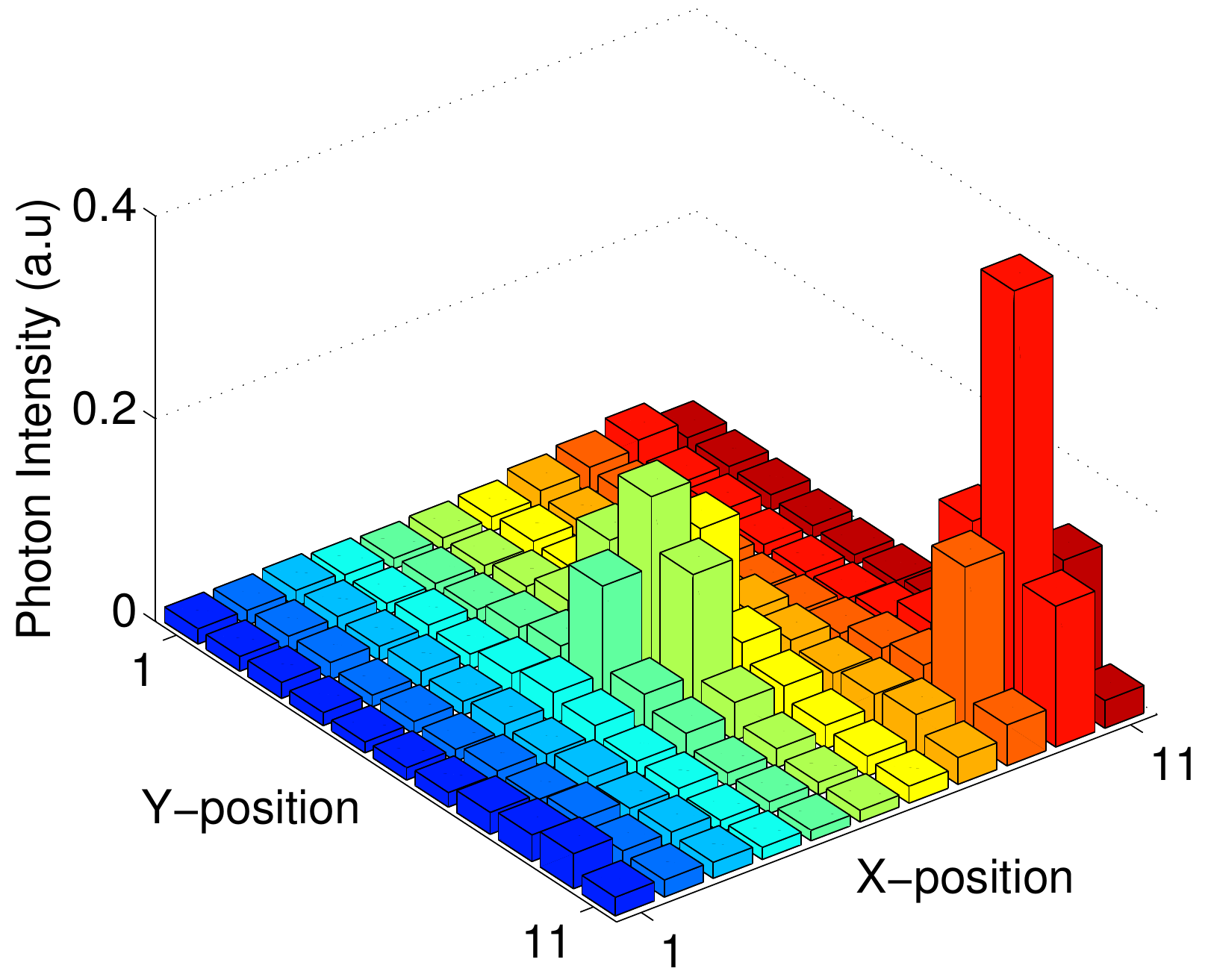}
\end{center}
\caption{Intensity distribution (in the physical basis, coin space traced over) after 24 steps $ (\approx \sqrt{N\, \mbox{ln} N} )$ of the driven QW. }\label{dri_grov_final_phys}
\end{figure}

We now apply the driven DTQW to the above scheme by examining a two-dimensional ($N=11\times11$) lattice with two vertices marked with the same coin, one whose position is known (at x=y=6), called the central vertex. Another vertex, called the target vertex, is randomly positioned on the lattice (here we use x=y=10 for clarity, but other positions yield similar results). We will then pump walkers into the central vertex with a uniform distribution over all coin states (Left, Right, Up, Down) and with a time-dependent phase that corresponds to the correct eigenfrequency ($\omega=0$) added at each time step. We then need to run the driven QW for a minimum number of steps that depends upon the spectral gap between our eigenmode of interest and the closest eigenmode in frequency. For a 2D lattice this number of steps is $\sqrt{N \mbox{ln} N}= 11\,\sqrt{2\mbox{ln}(11)} \approx 24$.

In figure~\ref{eigen_vector_grover} we plot the distribution of the eigenvector of interest in the physical basis (tracing over the coin states), which is seen to be localized at the two marked vertices. Figure \ref{ddtqw_grover_evo} shows the dynamics of the walker intensity during the DDTQW, in both the eigenbasis (dotted lines) and the physical basis (solid lines). The eigenvector of interest (Dotted, red line) is phase-matched and the closest eigenvectors in frequency (Dotted, blue line) are non-phase-matched. The walker intensity at the central vertex where the walkers enter is the solid red line and the intensity at the target mode is the solid blue line. Figure \ref{dri_grov_final_phys} shows the intensity distribution of walkers after 24 steps, where it can be seen that the two marked vertices have significantly different weight than other vertices, apart from the immediately surrounding ones. After more steps the walkers continue to accumulate in the marked vertices and the intensity distribution will tend to the eigenvector distribution. The advantages of this scheme is that it requires no extended initial state and also the walkers do not oscillate in-and-out of the marked vertex, thus we can measure at any time, after the initial waiting period. However, like another experimental scheme \cite{Bohm:2015p12316}, we have no reduction in the number of vertices needed so there is no resource advantage over the Grover search.\\
\indent In the next section we describe an experimental implementation of the driven QW on the one-dimensional line and the two-dimensional lattice.

\section{Proposal of Experimental Implementation}

We will discuss an experimental setup that could be used to test the predictions of these models, giving a detailed explanation for a possible realisation in one dimension. This is followed by a brief description of the necessary modifications for a two-dimensional extension of the walk with emphasis on the search algorithm.

\subsection{1-D Implementation}

\begin{figure}[ht!]
\begin{center}
\includegraphics[width=.75\columnwidth]{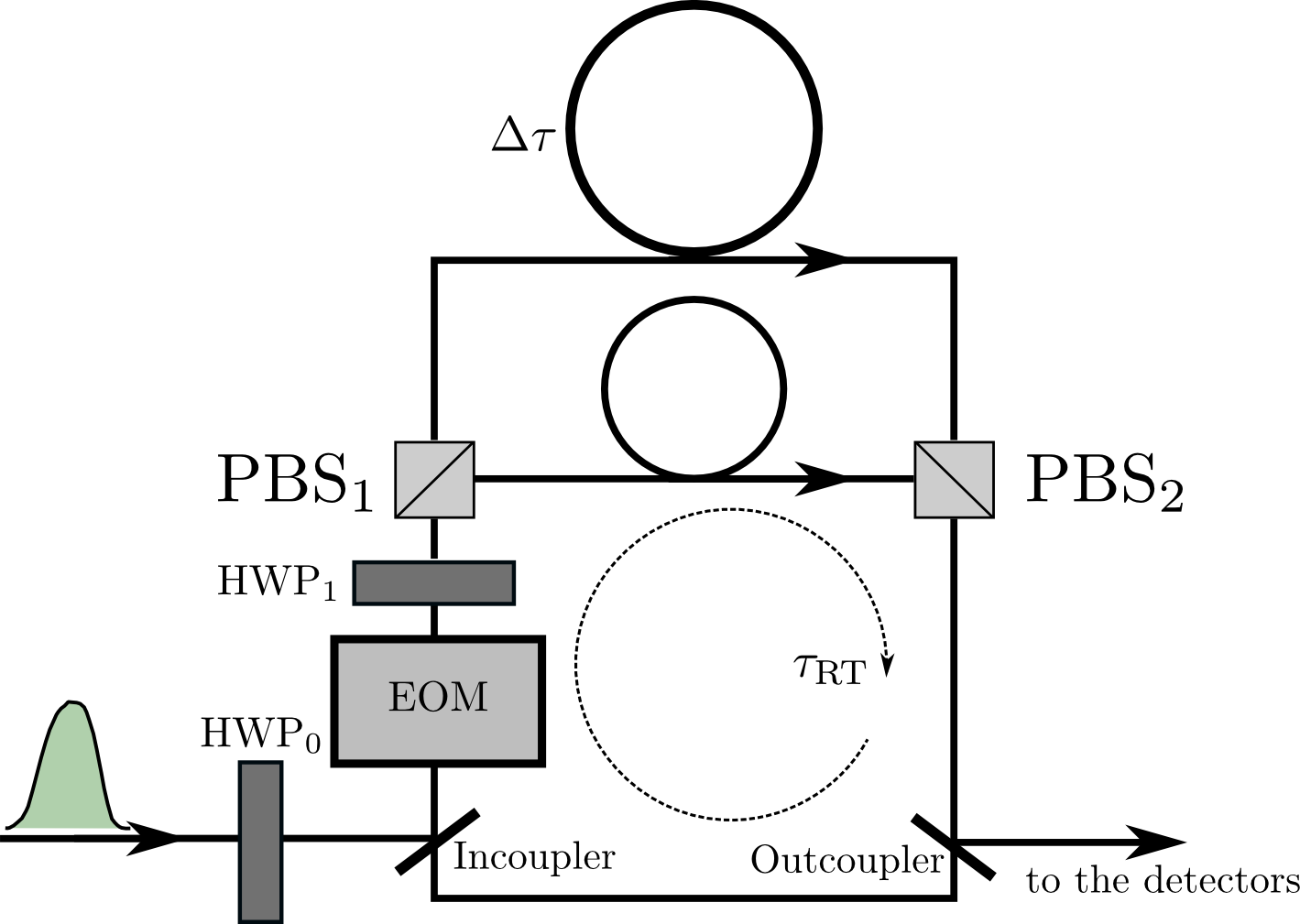}
\end{center}
\caption{Experimental setup of the DTQW as used in \cite{Schreiber:2010p5461} with half wave plate HWP$_1$ and EOM realising the (dynamic) coin operation, fibres of different lengths defining the round trip time $\tau_\mathrm{RT}$ and the position separation $\Delta\tau$. For details see text. \label{fig:QW_loop}}
\end{figure}

The implementation makes use of the time multiplexing-scheme of Schreiber {\it et.~al~} \cite{Schreiber:2010p5461} in an optical time-delay loop, see figure \ref{fig:QW_loop} for a simple diagram. The principle of time multiplexing is based on fibre loop delays used for mapping the position information of the walker onto the time domain. Here a weak coherent pulse plays the role of the walker whose polarization acts as the internal degree of freedom for the coin space. The initial pulse whose polarization can be set by standard linear elements (HWP$_0$) in front of the setup enters the loop at a partially reflective mirror, the incoupler. The coin operation is then realised by a wave plate inside the loop (HWP$_1$) or an electro-optic modulator (EOM) fast enough to address each position separately.
The EOM was previously used to introduce coin noise \cite{Schreiber:2010p5461} and to remove edges in the underlying graph to realise percolated \cite{Elster:2015p12200} and finite walks \cite{nits16arx}. For the Grover search scheme the EOM will be used to generate the special coins for the marked vertices.
After the polarization rotation the pulse is split at a polarizing beam splitter (PBS$_1$) and routed through fibres of different length such that the horizontal component obtains a delay of $\Delta \tau$ compared to the vertical component which realises the step operation according to (\ref{eq:QW_unitary}).
Before the pulses are fed back in the QW-loop a small portion of the light is coupled out to the detectors to measure the time evolution of the walker's wave function.
For detection another PBS and two avalanche photo diodes with a high dynamic range are used offering polarization resolved measurements and reliable intensity information.
In this scheme only pulses that have travelled exactly the same path result in the same time bin and accordingly will interfere with each other. This proves the setup to possess a high robustness against dephasing, a homogeneity over high step numbers and to be very resource efficient at the same time.

\begin{figure}[phtb]
\begin{center}
\includegraphics[width=.55\columnwidth]{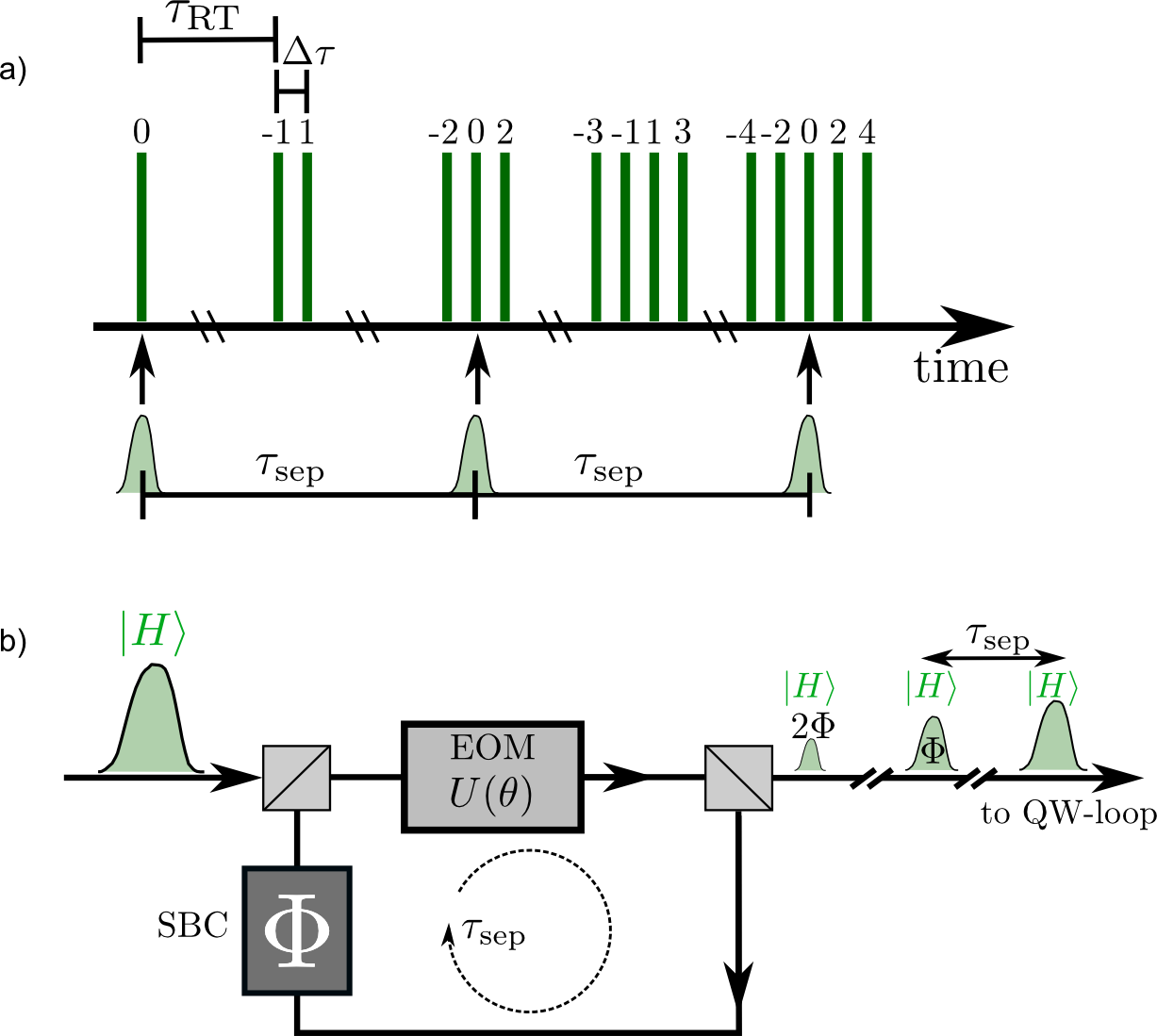}
\end{center}
\caption{a) Principle of time-multiplexing, i.e. the mapping of positions into the time domain (not to scale) using two different time-scales $\tau_{RT}$ and $\Delta \tau$. The timings of the new walkers to drive the walk at position $0$ are marked by arrows. b) The external loop to drive the QW: From an intense initial pulse a pulse train with a certain phase relation between pulses, matching intensities and a time separation of $\tau_\mathrm{sep} = 2\tau_{RT}+\Delta \tau$ is produced. For details see text.\label{fig:TM}}
\end{figure}

In a driven DTQW a new pulse will be added to the loop at every second time step (see marked positions in figure \ref{fig:TM}a) such that it interferes with the pulses already in the setup (pulses entering on odd or even step number have different parity and will not interfere with one another).
To achieve this we would use an additional external loop to produce the pulse train of the walkers with the desired phase and intensity properties (see figure \ref{fig:TM}b). 
The critical aspects to this setup are the timing of new pulses entering the system, the relative phase and amplitude of each new pulse.

First, the separation time $\tau_\mathrm{sep}$ of the new walkers must be adapted to the round trip times $\tau_{RT}$ of the setup, i.e. $\tau_\mathrm{sep} = 2\tau_{RT}+\Delta \tau$ for starting the new walkers always at position $0$ (see figure \ref{fig:TM}a) using appropriate fibres and potentially a delay stage for the fine tunings.

Next, we have to add a relative phase which the pulses in the external loop (and thus the newly-added pulses) will gain after each round trip compared to the pulses of light in the QW-loop. 
A Soleil-Babinet-Compensator (SBC) adds a fixed phase $\Phi$ to the remaining pulse in each round trip necessary for the phase-matching as described in Section \ref{sec_phase_match} and the experimental phase $\phi_\mathrm{exp}$ which the pulses in the QW-loop acquire during two steps.
Only then the essential coherence between the new and old pulses is given to ensure their interference capability.

Third, the intensity of the new pulses has to be considered:
when a horizontally polarised laser pulse enters the external loop it first passes an EOM, which represents the operator $\hat{U}(\theta)$ (for details see \cite{Elster:2015p12200})
\be
\hat{U}(\theta) = \left ( \begin{array}{cc} \cos(\theta) & i\sin(\theta) \\ i\sin(\theta)& \cos(\theta) \end{array} \right )~.
\ee
This rotates a part of the horizontally (H) polarized light into vertically (V) polarised light and at the second PBS the remaining H-light leaves the preparation loop whereas the V-light stays within.
The angle $\theta$ of the EOM defines the splitting ratio and thereby the intensity of the new walkers.
The pulses in the setup experience a loss per round trip (typically $l \approx 50\%$) which reduces their intensity.
Accordingly, each time a walker encounters $\hat{U}$, $\theta$ has to be adapted to regulate its intensity matching to those pulses which have already suffered from the losses in the QW-loop.
Of course the initial intensity of the laser pulse must be high enough to contain sufficient energy for the desired number of walkers.

By this means we have now prepared a coherent pulse train to drive the QW, which can now -- after setting its initial polarization (with e.g HWP$_0$ in figure \ref{fig:QW_loop})
- be fed into the QW-loop.

\subsection{2-Dimensional Quantum Walk and Search}

\begin{figure}[b]
\begin{center}
\includegraphics[width=.35\columnwidth]{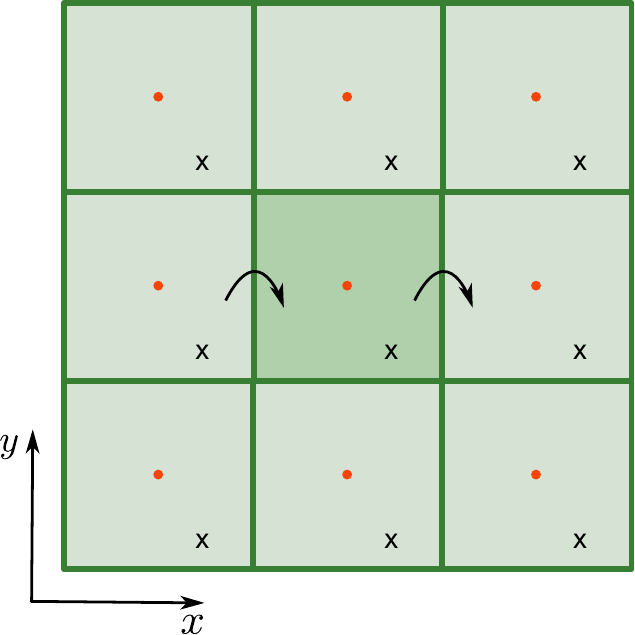}
\end{center}
\caption{Implementing periodic boundary conditions using copies of the original lattice (dark green region). The copies include the driving position (red dots) as well as the marked vertex (black crosses). The arrows indicate a pulse that leaves the original lattice to the right, when at the same time its copy enters it from the left imitating periodic boundaries.
 \label{fig:periodicBound}}
\end{figure}

The QW loop described above can be extended to a second dimension as demonstrated in \cite{Schreiber:2012p5592} with adjustable position-dependent coin operations. A driven 2D QW based on this experimental setup is realisable by adapting the timings and splitting ratios in the preparation loop accordingly. However, when implementing a search algorithm periodic boundary conditions are essential, which is typically hard to realise in an experiment. For walks of length N it is possible to imitate periodic boundary conditions by placing copies of the original lattice (including the marked vertex and the central, driven vertex) around it, see Fig. \ref{fig:periodicBound}.
That means that the pulses which leave the main region (dark green) at one side have a copy which enters from the opposite side at the same time. Now the QW must be driven at several positions of the extended lattice (red dots in Fig. \ref{fig:periodicBound}) at each step. This can be achieved by including a second loop in the preparation setup. It introduces a second time scale which can be matched to the position separation of the initial positions.

\section{Conclusions}

In this paper we have extended our recent, experimentally motivated, work in the area of continuous-time QW's with multiple walkers to the discrete-time case, describing what we term a driven DTQW. In this process the walkers, which are pulses of light, are coherently added at each time step of the walk, interfering with walkers already present. We analysed the dynamics in the eigenmode picture, as it was straight-forward to arrange the operators in a form that resembled the traditional QW dynamics. These dynamics relied on the product of displacement operators of the eigenmodes and we illustrated how eigenmodes of the system can either be phase-matched or not. Next, we demonstrated these dynamics with a numerical simulation of such a driven walk on a finite one-dimensional line with reflecting boundary conditions. Following this, we introduced a driven search algorithm based on the QW version of Grover's search. When the walkers enter with the correct phase, one that corresponds to an eigenmode that is predominantly located on the two marked vertices, the location of the unknown vertex can be determined. This new type of search has some advantages over the traditional Grover search as the dynamics do not let the walkers oscillate in-and-out of the marked vertices and no complex initial state is required. Finally, we have proposed how existing experimental implementations of QW's using optical time-delay loops could be used to simulate the driven scheme presented here, both for a one-dimensional line topology and a  two-dimensional lattice setup with periodic boundary conditions. 

\section{Acknowledgements}
C.S.H. and I.J. received financial support from Grants No. RVO 68407700 and No. GA\v{C}R 13-33906 S. 
L.S. and C.S. received funding from the European UnionÕs Horizon 2020 research and innovation program under the QUCHIP project GA no. 641039. S.B. and C.S. acknowledge funding by the DFG (Deutsche Forschungsgemeinschaft) via the Gottfried Wilhelm Leibniz-Preis.

\section{References} 
\bibliographystyle{unsrt}
\bibliography{driven_DTQW.bib}

\end{document}